\documentclass[aps,prl,twocolumn,superscriptaddress]{revtex4-2}

\usepackage{lineno}
\usepackage{amsthm}
\usepackage{amsmath}
\usepackage{amssymb}
\usepackage{mathrsfs}
\usepackage{txfonts}
\usepackage{graphics}
\usepackage{color} 
\usepackage{enumitem}
\usepackage{bm} 

\begin{document}

\preprint{}

\title{Topological unidirectional guided resonances emerged from interband coupling}


\author{Xuefan Yin}
\affiliation{Department of Electronic Science and Engineering, Kyoto University, Kyoto-Daigaku-Katsura, Nishikyo-ku, Kyoto 615-8510, Japan}
\author{Takuya Inoue}
\affiliation{Department of Electronic Science and Engineering, Kyoto University, Kyoto-Daigaku-Katsura, Nishikyo-ku, Kyoto 615-8510, Japan}
\author{Chao Peng}
\email{pengchao@pku.edu.cn}
\affiliation{State Key Laboratory of Advanced Optical Communication Systems and Networks, School of
Electronics, $\&$ Frontiers Science Center for Nano-optoelectronics,
Peking University, Beijing, 100871, China}
\affiliation{Peng Cheng Laboratory, Shenzhen 518055, China}

\author{Susumu Noda}
\email{snoda@qoe.kuee.kyoto-u.ac.jp}
\affiliation{Department of Electronic Science and
Engineering, Kyoto University, Kyoto-Daigaku-Katsura, Nishikyo-ku,
Kyoto 615-8510, Japan}


\date{\today}

\begin{abstract}
Unidirectional guided resonances (UGRs) are optical modes in photonic crystal (PhC) slabs that radiate towards one side without the need for mirrors on the other, represented from a topological perspective by the merged points of paired, single-sided, half-integer topological charges. In this work, we report a mechanism to realize UGRs by tuning the interband coupling effect originating from up-down symmetry breaking. We theoretically demonstrate that a type of polarization singularity, the circular-polarized states (CPs), emerge from trivial polarization fields owing to the hybridization of two unperturbed states. By tuning structural parameters, two half-charges carried by CPs evolve in momentum space and merge to create UGRs. Our findings show that UGRs are ubiquitous in PhC slabs, and can systematically be found from our method, thus paving the way to new possibilities of light manipulation.

\end{abstract}


\maketitle


Unidirectional emission is of fundamental interest in research fields including non-Hermitian physics \cite{bergholtz2021exceptional,leykam2017edge,shen2018topological} and singular optics \cite{dennis2009singular,Soskin_2016,gbur_singular_2016}, and can benefit many realistic applications such as on-chip lasers \cite{streifer1976analysis1,meier1999laser,imada1999coherent,Matsubara_GaN_2008,hirose2014watt,yoshida_double-lattice_2019,sakata_dually_2020,morita2021photonic} and energy-efficient grating couplers \cite{Wang_compact_slant_grating,Roelkens:07,Mekis_coupler_2011,markwade_grating_coupler,sun2015single,michaels2018inverse}. While most existing methods use mirrors made of metals or photonic-bandgap materials \cite{Roncone93_grating_coupler,Taillaert:04,Zhu:17}, or by utilizing the non-resonant blazing effect \cite{streifer1976analysis2,Miller:97} to forbid the radiation of light towards unnecessary ports, recent findings of unidirectional guided resonances (UGRs) \cite{Zhou_2016_single_side,yin_observation_2020} revealed that an eigenstate itself can radiate towards only a single side of the photonic crystal (PhC) slab without the need for a mirror on the other. From the view of topological photonics \cite{lu_topological_2014,khanikaev_two-dimensional_2017,ozawa_topological_2019,wang_topological_2020}, the UGRs were connected to polarization singularities \cite{nye1983lines,Flossmann_polarization_2008,thomas_observation_2015,F_sel_2017,Bliokh_2019,chen2019singularities,shilei_polarization_singularity_2021,yuri_singularity_review_2021,wang_polarization_2021,yuri_topological_2021} in momentum space, represented by topological charges \cite{zhen_topological_2014,bulgakov2017bound,shilei_observation_2018,alu_experimental_2018,liuwei_global_charge,zeng_dynamic_2021}: they are the merged $V$ points (vortex center of
polarization fields) from paired $C$ points (circular-polarized states) that carry the same signed half-integer topological charges on a single side. As reported, such half-charges can originate from splitting an integer charge carried by a bound state in the continuum (BIC) \cite{von_neuman_uber_1929,marinica2008bound,zhen_topological_2014,hsu_bound_2016} as $q=1 \rightarrow 1/2+1/2$ \cite{shilei_circular,yin_observation_2020,ye_singular_2020,yoda_generation_2020}, or they can be spawned from ``the void'' as $q=0 \rightarrow 1/2+(-1/2)$ \cite{zeng_dynamic_2021}. Both cases obey the conservation law of topological charges \cite{zhen_topological_2014,hsu_bound_2016,liuwei_global_charge}. 

From such a topological view, the key to realizing UGRs is to create paired circular-polarized states (CPs). Although splitting from an integer charge of a BIC has been shown as a fairly intuitive example, it is still unclear how to deterministically separate and merge the CPs from the trivial polarization field, namely the void. Considering that the UGRs intrinsically do not depend on any symmetry, their realization is a formidable engineering challenge, especially when no appropriate BICs can be found as a starting point.

\begin{figure}[htbp] 
 \centering 
 \includegraphics{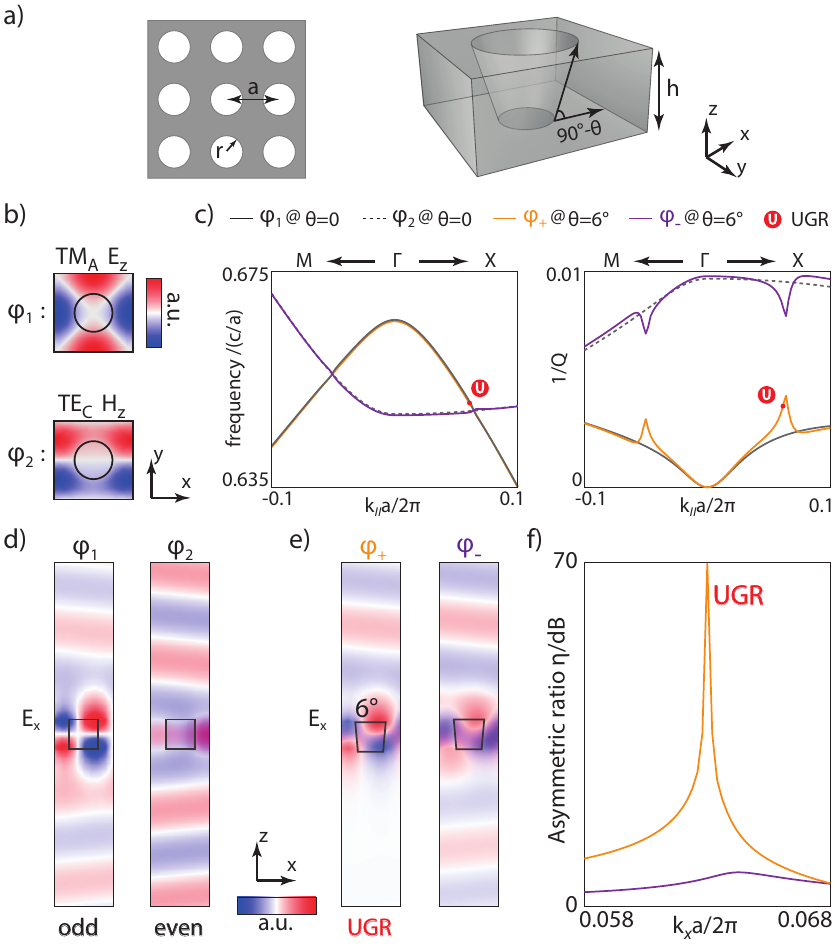} 
\caption{UGR raised by interband coupling. a, schematic of a PhC slab, with slab thickness $h/a=0.48$, radius $r/a=0.238$. b, profiles of TM$_A$ ($\varphi_1$) and TE$_C$ ($\varphi_2$) modes. c, band structures of original resonances $\varphi_{1,2}$ (gray and gray dashed lines) and perturbed resonances $\varphi_{+,-}$ (yellow and purple lines) along the M-$\Gamma$-X direction,
with a UGR at $(k_x=0.063,k_y=0)$ in $\varphi_+$. d,e, profiles of $\varphi_{1,2}$ and $\varphi_{+,-}$ at $(k_x=0.063,k_y=0)$. f, asymmetric radiation ratio $\eta$ of $\varphi_{+,-}$.}
\label{Fig1} 
\end{figure}
In this letter, we report a mechanism to realize UGRs without the premise of BICs. Specifically, we find that the interband couplings raised from up-down mirror symmetry breaking can be utilized as an effective degree of freedom (DOF) to hybridize the orthogonal bands in an unperturbed system. As a result of interband coupling, the paired CPs with opposite half-charge of $q=\pm 1/2$ are spawned from the void of $q=0$ and evolve in momentum space by adjusting the interband coupling strength. 
In the case of a system also possessing in-plane mirror symmetry, a pair of mirror-positioned half-charges with the same sign in single-sided radiation can merge at the high-symmetry lines in momentum space, thus generating UGRs.

To be more specific, we focus on a two-dimensional (2D) square-latticed PhC slab with symmetric upper and lower claddings. We break the up-down mirror symmetry of the PhC slab by isotropically tilting the sidewalls of the air holes. Consequently, interband coupling occurs between two crossing bands which are originally orthogonal to each other. When the interband coupling is strong enough, it could give rise to two perturbed eigenstates with nontrivial topology upon the radiation: a CP pair with oppositely-signed half-charges. 
By tuning the tilting angle, the half-charges continuously move and merge in the $\Gamma-X$ or $\Gamma-M$ directions, creating multiple UGRs. Our investigation shows that the UGRs are even more ubiquitous than expected, because band-crossing and corresponding interband coupling are quite common in PhC slabs regardless of their specific material and geometry, such as refractive-index contrast, slab thickness, and hole shape.


We start from a free-standing Si$_3$N$_4$ slab patterned with square-lattice air holes. The sidewalls of the air holes can be either perfectly vertical or isotropically tilted, measured by an angle  $\theta$ (Fig.~\ref{Fig1}a). According to Bloch's theorem, the eigenstates are in the form of  $U_{x,y,z}=E_{x,y,z}(\bm{r})e^{-i\bm{k}_{\varparallel}\cdot\bm{r}}e^{i\omega t}$ with in-plane wavevector $\bm{k}_{\varparallel}=(k_x,k_y)\beta_0$ characterizing $k_\varparallel$-momentum space, where $\beta_0=2\pi/a$ and $a$ is the lattice constant. For vertical sidewalls, we focus on the TM$_A$ mode (denoted as $\varphi_1$) and TE$_C$ mode (denoted as $\varphi_2$) around the 2nd-$\Gamma$ point (Fig.~\ref{Fig1}b). Note that in the band structures (Fig.~\ref{Fig1}c, gray lines and gray dashed lines), $\varphi_1$ and $\varphi_2$ cross at $(k_x=0.064,k_y=0)$ and $(k_x=0.035,k_y=0.035)$, respectively. Due to the up-down mirror symmetry, both $\varphi_1$ and $\varphi_2$ are eigenfunctions of parity operator $\mathbf{\hat{P_z}}$, but with opposite eigenvalues: $\varphi_1$ is odd with $\sigma_z=-1$; $\varphi_2$ is even with $\sigma_z=1$ (Fig.~\ref{Fig1}c). Therefore, the orthogonality forbids any interband coupling between them during the band crossing. 

Then, we slightly tilt the sidewalls, turning the cylindrical air holes into truncated cones with $\theta=6^\circ$. Owing to up-down mirror symmetry breaking, $\varphi_1$ and $\varphi_2$ couple and give rise to two perturbed eigenstates $\varphi_+$ and $\varphi_-$ \cite{tanaka_theoretical_2003}. As shown in Fig.~\ref{Fig1}c, the real parts of their frequencies cross while imaginary parts anti-cross. As confirmed in Fig.~\ref{Fig1}e, $\varphi_{+,-}$ no longer preserve definite $z$-parity, and thus the radiation becomes up-down asymmetric, characterized by an asymmetric radiation ratio $\eta=\gamma_t/\gamma_b$, where $\gamma_{t,b}$ are decay rates towards the top and bottom. A UGR is found at $(k_x=0.063,k_y=0)$ upon $\varphi_+$ band near one of the crossing points, with an asymmetric ratio of $70$ dB (Fig.~\ref{Fig1}h).

To investigate the details of interband coupling, we derive a two-level model from perturbation theory. Specifically, we start from a up-down symmetric system, in which $\varphi_1$ and $\varphi_2$ are two eigenstates of the unperturbed Hamiltonian $\hat{\mathbf{H}}_0=1/\varepsilon_0(x,y)\nabla\times\nabla\times$ with eigenvalues $\lambda_{1,2}$ (other eigenstates are neglected). With the tilting, a perturbation $\Delta\hat{\mathbf{H}}$ is added as $
\hat{\mathbf{H}}=\hat{\mathbf{H}}_0+\Delta\hat{\mathbf{H}}$. The detailed formulation of $\hat{\mathbf{H}}$ and $\Delta\hat{\mathbf{H}}$ can be found in Supplementary Section 1.

Accordingly, the unperturbed eigenstates $\varphi_{1,2}$ can be utilized as a set of bases of subspace $(\varphi_1,\varphi_2)$ to depict the perturbed Hamiltonian $\hat{\mathbf{H}}$ in a form of a 2$\times$2 matrix:
\begin{eqnarray}
\label{eq:1}
\mathcal{H}=\left[\begin{array}{cc}
     \left<\varphi_1\mid\Delta\hat{\mathbf{H}}\mid\varphi_1\right>+\lambda_1&\left<\varphi_1\mid\Delta\hat{\mathbf{H}}\mid\varphi_2\right>  \\
     \left<\varphi_2\mid\Delta\hat{\mathbf{H}}\mid\varphi_1\right>&\left<\varphi_2\mid\Delta\hat{\mathbf{H}}\mid\varphi_2\right>+\lambda_2
\end{array}\right]
\end{eqnarray}
and the hybrid eigenstates can be written as
\begin{eqnarray}
\label{eq:2}
\varphi_{+,-}(\bm{k}_\varparallel)=a_{+,-}(\bm{k}_\varparallel)\varphi_1(\bm{k}_\varparallel)+b_{+,-}(\bm{k}_\varparallel)\varphi_2(\bm{k}_\varparallel)
\end{eqnarray}
with the superposition coefficients given by the eigenvectors $[a,b]^T_{+,-}$ of $\mathcal{H}$. According to Eq.~S6, the perturbation $\Delta\hat{\mathbf{H}}$ is proportional to the odd function $\tan\theta\cdot z$, and thus the inner-product terms in matrix $\mathcal{H}$ can be simplified as:
\begin{eqnarray}
\label{eq:3}
 \left<\varphi_{j}\mid\Delta\hat{\mathbf{H}}\mid\varphi_{k}\right>\propto\left<\varphi_{j}\mid z \mid\varphi_{k}\right>\tan\theta=\kappa_{jk}\tan\theta
\end{eqnarray}
As a result, only the eigenstates with opposite $z$-parity can couple to each other through the anti-diagonal coupling terms. As illustrated in Fig.~\ref{Fig1}d, $\varphi_{1,2}$ fulfill this prerequisite.

\begin{figure}[htbp] 
 \centering 
 \includegraphics{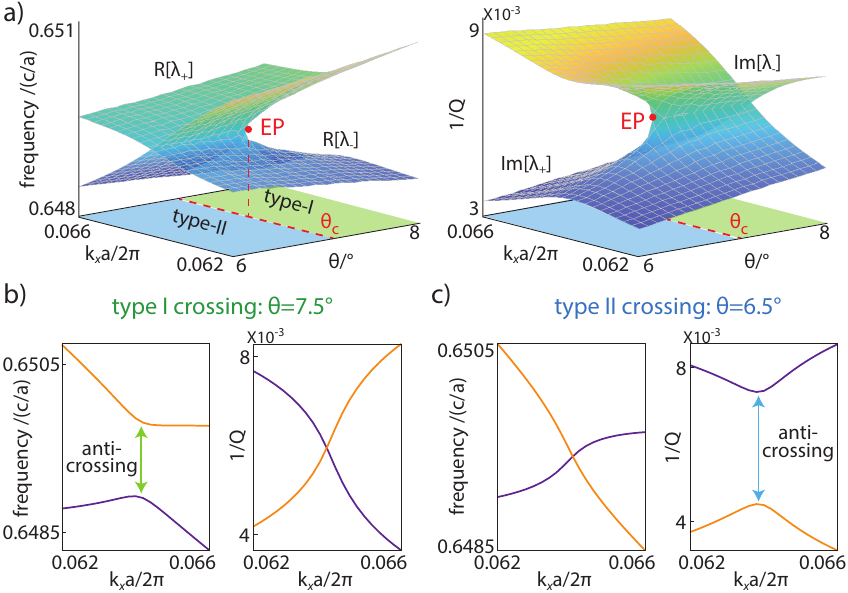} 
\caption{Interband coupling scenario. a, band structures of perturbed resonances $\varphi_{+,-}$ in parameter space $(\theta, k_x)$ with $k_y$ fixed to zero. Red dot denotes EP. b,c, examples of type-I and type-II crossings when $\theta=7.5^\circ/6.5^\circ$.}
\label{Fig2} 
\end{figure}

Accordingly, the eigenvalues of $\mathcal{H}$ can be derived as:
\begin{eqnarray}
\label{eq:4}
\lambda_{+,-}=\frac{\lambda_1+\lambda_2}{2}
\pm\sqrt{\frac{(\lambda_1-\lambda_2)^2}{4}+\kappa_{12}\kappa_{21}\tan\theta^2}
\end{eqnarray}. Obviously, there exists a critical angle $\theta_c$ with proper $\bm{k}_\varparallel$ for which $(\lambda_1-\lambda_2)^2/4+\kappa_{12}\kappa_{21}\tan{\theta_c}^2=0$, namely at which the complex eigenvalues $\lambda_{+,-}$ are degenerate. Such points are known as the exceptional points (EPs) \cite{heiss2012physics,dembowski2001experimental,zhen2015spawning,doppler2016dynamically,zhou_observation_2018,miri2019exceptional,bergholtz2021exceptional}: the complex crossing points of eigenvalues in parameter space $(\theta,\bm{k}_\varparallel)$. We obtain $\theta_c=6.9^\circ$ along the $\Gamma$-$X$ direction ($\bm{k}_\varparallel=k_x$) for the PhC shown in Fig.~\ref{Fig1}. We further plot the complex band structures of $\varphi_{+,-}$ in parameter space $(\theta,k_x)$, as presented in Fig.~\ref{Fig2}a. An isolated EP is found  at $(\theta=6.9^\circ,k_x=0.064)$.

As marked in Fig.~\ref{Fig2}a, the critical angle $\theta_c$ divides the parameter space $(\theta,k_x)$ into two different regions \cite{keck2003unfolding}:
 \begin{description}
    \item [Type-I crossing] for $\theta>\theta_c$, the real parts of $\lambda_{+,-}$ anti-cross and the imaginary parts cross (green region).
    \item[Type-II crossing] for $\theta<\theta_c$, the real parts of $\lambda_{+,-}$ cross and the imaginary parts anti-cross (blue region).
    \end{description}

By continuously increasing $\theta$ from zero, the crossing type first belongs to type-II, and then transitions to type-I when passing by the EP at $\theta_c=6.9^\circ$. In fact, similar transitions can be found not only along the $\Gamma$-$X$ direction, but also along arbitrary $\bm{k}_\varparallel$ directions in the Brillouin zone (BZ) due to $C_4$ symmetry. In Supplementary Section 2, we present the interband coupling scenario along $k_y=k_x/2$ direction as an example, for which the EP resides at $(\theta_c=7.15^\circ,k_x=0.0517)$.

We further elaborate on the possibilities of creating polarization singularities from interband coupling. Eq.~\ref{eq:2} shows that $\varphi_{+,-}$ are hybridized from $\varphi_{1,2}$. Accordingly, the far-field polarization of $\varphi_{+,-}$ also correlate to the polarization of $\varphi_{1,2}$ denoted by $d^{s}_{1,2}=c_{y;1,2}^{s}/c_{x;1,2}^{s}$. Here $c^{s}_{x,y;1,2}$ are the complex radiation amplitudes of $\varphi_{1,2}$ towards the top and bottom sides, marked by superscripts $s \in \{t,b\}$. From the two-level model, the polarization of  $\varphi_{+,-}$ can be derived as: 
\begin{eqnarray}
\label{eq:5}
d^{s}_{+,-}=m^{s}_{+,-}d^{s}_1+(1-m^{s}_{+,-})d^{s}_2, ~~~~ s \in \{t,b\}
\end{eqnarray}
where $m_{+,-}^{s}$ are complex coefficients representing the interband coupling (a detailed discussion is provided in Supplementary Section 3). Obviously, the intrinsic radiation characteristics of perturbed eigenstates $\varphi_{+,-}$ are modified by the interband coupling effect. Once the polarization $d^{s}_{1,2}$ are linearly independent, polarization $d^{s}_{+,-}$ can be arbitrary values in principle if appropriate  $m^{s}_{+,-}$ are applied. 

\begin{figure}[htbp] 
 \centering 
 \includegraphics{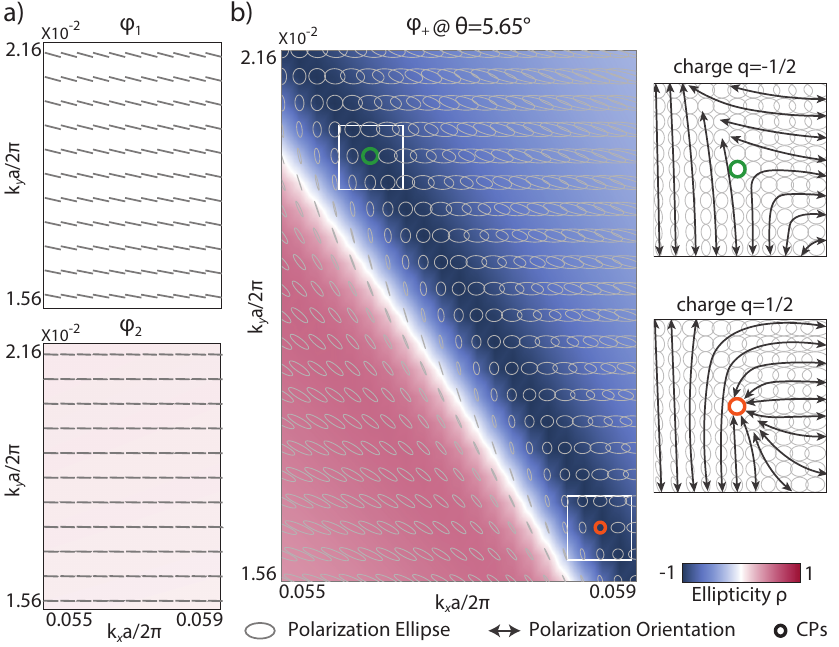} 
\caption{CPs carrying half-charges emerged from interband coupling. a, downward polarization and ellipticity $\rho$ of $\varphi_{1,2}$. b, left panel: downward polarization and ellipticity of perturbed resonance $\varphi_+$ when $\theta=5.65^\circ$, where two CPs (green and orange circles) with ellipticity of $-1$ emerge. right panels: zoom-in windings of polarization orientation around the CPs.}
\label{Fig3} 
\end{figure}

Here we focus on a type of polarization singularity: CPs. Without loss of generality, we focus on the downward radiation of $\varphi_+$. The condition of CP becomes $d^b_+=m^b_+(d^b_1-d^b_2)+d^b_2=\pm i$. Clearly, two DOFs are necessary to fulfill this condition. In other words, the CPs can emerge in 2D parameter space, i.e. 2D $k_\varparallel$-momentum space, by optimizing complex coefficient $m_+^b$ (which also has 2 DOFs) from interband coupling even though $d^b_{1,2}$ themselves are trivial in polarization.

We plot the polarization ellipses and ellipticity $\rho$ (colored map) \cite{Mcmaster1954Polarization} of $d^{b}_{1,2}$ in Fig.~\ref{Fig3}a, showing that both $\varphi_1$ and $\varphi_2$ are independent but near-linearly polarized with $\rho \approx 0$, namely no polarization singularities such as CPs are found. Then by tilting the sidewalls, complex coefficient $m^b_{+}$ is induced from interband coupling, modifying the polarization of $\varphi_+$. For a small $\theta$, the coupling strength is not strong enough to create any CPs from trivial near-linear polarization, which we refer to as the void. When $\theta$ is increased to $\theta_v\sim5.65^\circ$, a pair of CPs emerges simultaneously at $(k_x=0.0554,k_y=0.0212)$ and $(k_x=0.0588,k_y=0.0159)$ (Fig.~\ref{Fig3}b). It is noteworthy that the ellipticity of both CPs are identical as $\rho=-1$ (dark blue), showing that they are both left-handed CP (LCP).

The windings of polarization orientation~\cite{zhen_topological_2014} in the vicinity of two CPs are plotted in the right panel of Fig.~\ref{Fig3}b, showing that the two CPs carry half-charges with opposite polarity ($q=\pm 1/2$). Recalling the fact that when interband coupling is absent or sufficiently weak, $\varphi_+$ has no polarization singularity ($q=0$), we conclude that the charge evolves as paired, opposite half-charges ($q=\pm 1/2$) carried by two LCPs spawned from the void ($q=0$) as $0 \rightarrow 1/2+(-1/2)$. Obviously, such an evolution obeys the conservation law of topological charges. 

Next, we investigate how to merge two half-charges to create the UGRs. Before discussing the details, it is worthwhile to discuss the symmetry of system. As explained previously, under up-down mirror symmetry breaking, the polarization singularity of CPs, which appears for downward radiation of $\varphi_+$, cannot coincide at the same $k_\varparallel$ point for upward radiation, and thus we can remain focused on $d_+^b(k_\varparallel)$. Owing to $C_4$ symmetry, a pair of opposite half-charges in one quadrant of momentum space would duplicate themselves to the other three quadrants, and evolve together under the constraint of $C_4$ symmetry. Notice that, to create UGRs, two half-charges with the same sign should be merged, which must come from different void points. Here, the in-plane mirror symmetries ensures paired mirror-positioned voids with respect to high-symmetry lines, providing the same signed half-charges pairs.

\begin{figure}[htbp] 
 \centering 
 \includegraphics{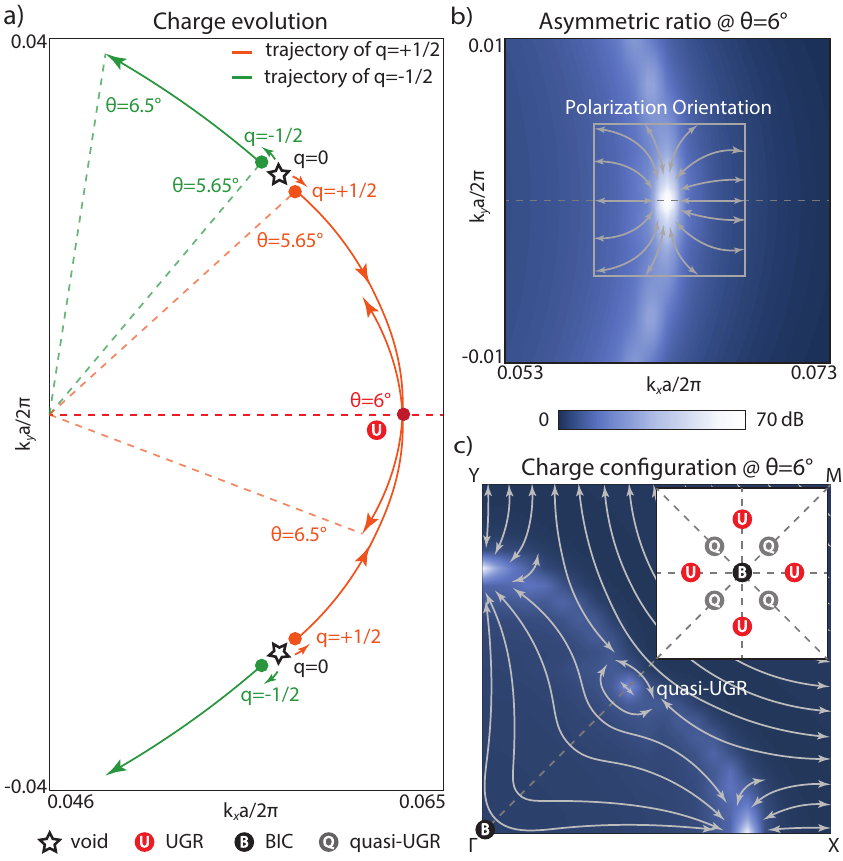} 
\caption{UGRs emerged from interband coupling. a, Evolution of half-charges in $d_+^b$ in right half of momentum space when $\theta$ is varied. b, Asymmetric radiation ratio (map) and the polarization orientation (arrows) around the UGR. c, Asymmetric radiation ratio (map) and the polarization orientation (arrows) in a quarter of momentum space. Inset: charge configuration in whole momentum space.}
\label{Fig4} 
\end{figure}

The detailed charge evolution of $d_+^b$ is illustrated in Fig.~\ref{Fig4}a. Here we plot the right-half of momentum space. Owing to $y$-mirror symmetry, two pairs of CPs are spawned from two mirror-positioned voids when $\theta>\theta_v=5.65^\circ$, and start to evolve from $(k_x=0.057,k_y=\pm 0.0186)$ (marked by stars), respectively. The CPs spawned from the same void have the same helicity but carry opposite half-charges. Recall that as a type of topological invariant, half-charges are robust in momentum space. By further increasing $\theta$, the negative half-charges ($q=-1/2$) depart towards the far-end of the $\pm k_y$ directions, respectively (green lines), while the positive half-charges ($q=+1/2$) which originate from two different voids move towards each other (orange lines). At $\theta=6^\circ$, the two positive half-charges meet on the $k_x$ axis at $(k_x=0.063,k_y=0)$, and merge into an integer topological charge as $ 1/2+1/2\rightarrow +1$ (red U marker). Consequently, the downward radiation of $\varphi_+$ is completely forbidden and an upward radiating UGR emerges. For $\theta$ exceeding $6^\circ$, the integer charge $q=+1$ splits again into a pair of half-charges as $+1 \rightarrow 1/2+1/2$. Then, the half-charges depart away from each other and the UGR disappears.

The map of asymmetric ratio $\eta$ for the UGR in $d_+^b$ is plotted in Fig.~\ref{Fig4}b. The ratio reaches $70$ dB at $(k_x=0.063,k_y=0)$, which confirms the existence of UGR. As shown in the inset of Fig.~\ref{Fig4}b, a vortex is clearly observed in the polarization orientation fields of downward radiation, giving rise to an integer topological charge of $q=+1$ as expected. 

The role of in-plane mirror symmetry in creating UGRs is clearly shown in Fig.~\ref{Fig4}a: the trajectories of half-charges spawned from two mirror-positioned voids are symmetric with respect to the $k_x$ axis, protected by $y$-mirror symmetry. Therefore, two positive half-charges steadily evolve toward each other and then merge at the $k_x$ axis, creating UGRs along the $\Gamma-X$ direction. Similarly, $x$-mirror symmetry protects the trajectories with respect to the $k_y$ axis, leading to the creation of UGRs along the $\Gamma-Y$ direction (Fig.~\ref{Fig4}c). Owing to $C_4$ symmetry, the $\Gamma-X$ and $\Gamma-Y$ directions are equivalent, and 4 UGRs emerge simultaneously in total as shown in the inset of Fig.~\ref{Fig4}c.

Besides the $\Gamma-X$ and $\Gamma-Y$ directions, $\Gamma-M$ directions are also high-symmetry lines owing to $C_4$ symmetry, and thus the UGRs can be similarly found from merging two positive half-charges spawned from paired mirror-positioned voids with respect to the $\Gamma-M$ axes. However, recall the fact that the UGR in $\varphi_+$ is a result by hybridizing $\varphi_1$ and $\varphi_2$. As shown in Fig.~\ref{Fig1}c, although the dispersion of $\varphi_1$ is fairly isotropic with respect to BZ center, the dispersion of $\varphi_2$ is considerably anisotropic. As a result, the condition of realizing UGRs in the $\Gamma-X$ and $\Gamma-M$ directions could be slightly different with respect to $\theta$. In other words, when two positive half-charges merge in the $\Gamma-X$ direction, another pair of positive half-charges are close to but do not exactly arrive at the high-symmetry line in the $\Gamma-M$ direction (Fig.~S3a). Consequently, the asymmetric radiation ratio is high but not infinite (Fig.~S3b). We refer this kind of state as a ``quasi-UGR''. A more detailed discussion can be found in Supplementary Section 4.

As shown in Fig.~\ref{Fig4}c, the UGRs emerge in the $\Gamma-X$ direction at $\theta=6^\circ$ with $\eta=70$ dB. For the same parameters, the highest asymmetric radiation ratio in the $\Gamma-M$ direction is $\eta=25.8$ dB (Fig.~S3b). As illustrated by the topological charge configuration (inset, Fig.~\ref{Fig4}c), there exist 1 BIC ($\Gamma$), 4 UGRs ($\Gamma$-$X$,$Y$), and 4 quasi-UGRs ($\Gamma$-$M$) in the BZ. Moreover, a high-asymmetric-radiation ring (colored map, Fig.~\ref{Fig4}c) emerges that connects those UGRs and quasi-UGRs. 

It is noteworthy that $C_4$ symmetry is crucial for multiple UGRs, but it is not necessary for creating UGRs in general. In Supplementary Section 5, we reduce the in-plane symmetry from $C_4$ to $y$-mirror only, by changing the circular air holes to triangular ones (Fig.~S5). As a result, a single isolated UGR is found at the $k_x$ axis of $(k_x=0.04,k_y=0)$ with $\eta=68.3$ dB. 

\begin{figure}[htbp] 
 \centering 
 \includegraphics{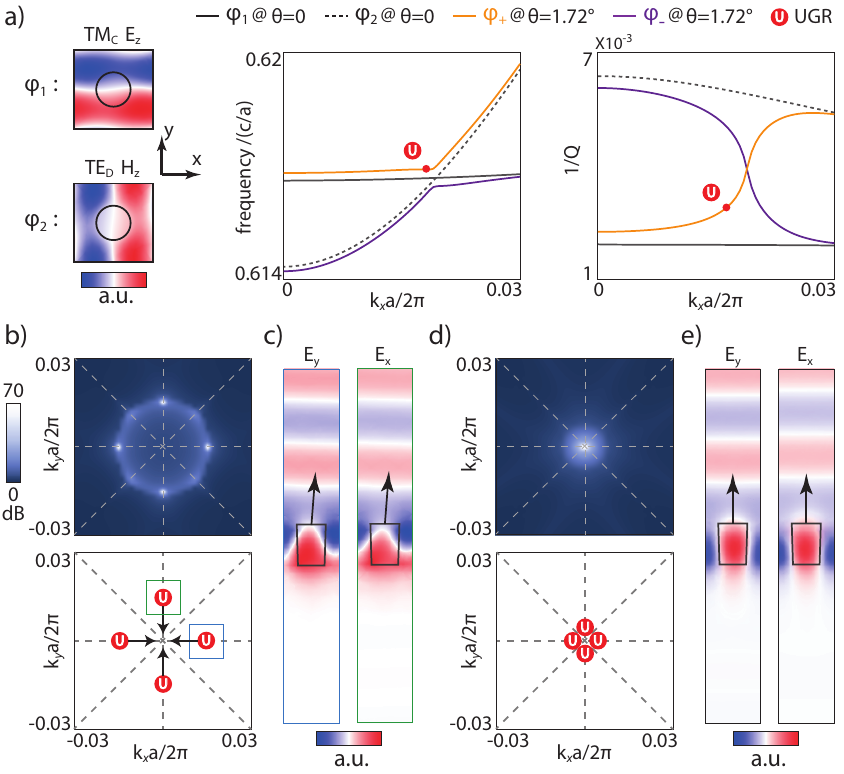} 
\caption{Merging UGRs. a, left panels: profiles of TM$_C$ ($\varphi_1$) and TE$_C$ ($\varphi_2$) modes. right panels: band structures of original resonances $\varphi_{1,2}$ (gray and gray dashed lines) and perturbed resonances $\varphi_{+,-}$ (yellow and purple lines), with a UGR at $(k_x=0.0172,k_y=0)$ in $\varphi_+$ when $h/a=0.65$, $r/a=0.219$ and $\theta=1.72^\circ$. b, asymmetric radiation ratio and charge configuration of $\varphi_+$. c, profiles of UGRs in $\varphi_+$ at $(k_x=0.0172,k_y=0)$ (blue box) and $(k_x=0,k_=0.0172)$ (green box), respectively. d,e, asymmetric radiation ratio, charge configuration and profiles of merging UGRs in $\varphi_+$ at $\Gamma$ point.}
\label{Fig5}
\end{figure}

Until now, we have focused on the TM$_A$ and TE$_C$ modes as unperturbed bases $(\varphi_1, \varphi_2)$ as shown in Fig.~\ref{Fig1}, and the band-crossing belongs to type-II when UGRs appear at $\theta=6^\circ$.
Given that any two modes with opposite $z$-parity are possible to couple, we turn to investigate another set of unperturbed bases, namely TM$_C$ mode (denoted as $\varphi_1$) and TE$_D$ mode (denoted as $\varphi_2$) near the 2nd-$\Gamma$ point (left panel, Fig.~\ref{Fig5}a). It is readily confirmed that these modes fulfill the parity requirement of interband coupling.

Specifically, at a tilting angle of $\theta=1.72^\circ$, type-I crossing can be identified from the complex band structures of Fig.~\ref{Fig5}a. An upward radiating UGR can be found upon the hybrid eigenstate $\varphi_+$ at $(k_x=0.0172,k_y=0)$. The asymmetric radiation ratio $\eta$ and charge configuration are presented in Fig.~\ref{Fig5}b. Clearly, 4 UGRs (red U markers) co-exist in the $\Gamma-X$ and $\Gamma-Y$ directions with $\eta$ exceeding $70$ dB, and together they form a high-asymmetric-radiation ring (upper panel of Fig.~\ref{Fig5}b). Further, we calculate the electric field profiles of two UGRs residing on the $+k_x$ and $+k_y$ axes (blue and green boxes in the lower panel of Fig.~\ref{Fig5}b) and plot them in Fig.~\ref{Fig5}c. Such UGRs are either $y$- or $x$-polarized, correspondingly. 


Protected by in-plane mirror symmetry, the UGRs can robustly exist and continuously evolve along the high-symmetry lines \cite{yin_observation_2020} (detailed discussions are provided in Supplementary Section 6). For instance, by tuning the slab thickness $h$ and hole radius $r$ while fixing the tilting angle $\theta=1.72^\circ$, the 4 off-$\Gamma$ UGRs in Fig.~\ref{Fig5}b move towards the BZ center in an isotropic manner. With $h/a \to 0.665$ and $r/a \to 0.217$, the UGRs merge at the $\Gamma$ point, and the high-asymmetric-radiation ring shrinks into a point, as shown in Fig.~\ref{Fig5}d. As a result, the unidirectional radiation becomes vertical for both $E_x$ and $E_y$ components (Fig.~\ref{Fig5}d).

It is noteworthy that, although the UGRs raised from interband coupling involve two unperturbed eigenstates, their unidirectional emission behavior is an intrinsic characteristics of one perturbed eigenstate, which does not originate from the interference between two independently excited resonances  \cite{suh2004temporal,chang2012high}. Furthermore, since the UGRs selectively close one of the radiation channels, they exhibit interesting and abnormal phenomena under external excitation \cite{Zhang_observation_2021}, leading to various applications for light manipulation.

In conclusion, we have proposed a systematic method to realize UGRs in PhC slab by coupling two unperturbed eigenstates from up-down mirror symmetry breaking. We find that the CPs carrying paired, opposite-signed topological half-integer charges are spawned from the void through interband coupling and can be controlled to merge only at one side of PhC slab, thus creating a UGR. Our findings shed light upon new possibilities for creating and utilizing polarization singularities in momentum space, providing a vivid picture for manipulating the optical radiation \cite{Yin_pr_2020} in various applications such as light trapping \cite{jin_topologically_2019,Kang_off_Gamma_merging_2021,CHEN2021}, generation of vortex beams \cite{shen_optical_2019,huang_ultrafast_2020,yang_spin-momentum-locked_2020,wang_generating_2020,zhang_tunable_2020} and polarization conversion \cite{guo2017topologically}.
 
\begin{acknowledgments}
The authors are grateful to Prof. B.S. Song, Dr. J. Gelleta, Dr. Z. Zhang and Dr. Y. Hu for helpful discussions. This work was partially supported by Grant-in-Aid for JSPS Research Fellow (21F20356). C. Peng was supported from National Natural Science Foundation of China (61922004 and 62135001), National Key Research and Development Program of China (2020YFB1806405) and Major Key Project of PCL (PCL2021A04 and PCL2021A14). X. Yin was supported by Research Fellowships of the Japan Society for Promotion of Science.
\end{acknowledgments}


\end{document}